\documentstyle[11pt,epsf]{article}
\newcommand{\postscript}[2]
   {\setlength{\epsfxsize}{#2\hsize}
   \centerline{\epsfbox{#1}}}
\newtheorem{theorem}{Theorem}
\newtheorem{corollary}{Corollary}
\newtheorem{definition}{Definition}
\newtheorem{assumption}{Assumption}
\newtheorem{lemma}{Lemma}
\newcommand{\x}{{\mathbf x}}
\newcommand{\ut}{\tilde{u}}
\newcommand{\cL}{{\cal L}}

\newcommand{\noi}{\noindent}
\newcommand{\bi}{\begin{itemize}}
\newcommand{\ei}{\end{itemize}}
\newcommand{\be}{\begin{equation}}
\newcommand{\ee}{\end{equation}}
\newcommand{\bea}{\begin{eqnarray}}
\newcommand{\eea}{\end{eqnarray}}
\newcommand{\ba}{\begin{array}}
\newcommand{\ea}{\end{array}}
\newcommand{\D}{\Delta}

\newcommand{\To}{\rightarrow}

\begin{document}
\def\theequation {\arabic{equation}}

\begin{center}
{\bf\Large Accelerated Imaginary-time Evolution Methods for the
Computation of Solitary Waves}

Jianke Yang and Taras I. Lakoba

{\small Department of Mathematics and Statistics, University of Vermont, Burlington, VT 05401}
\end{center}




Summary:

Two accelerated imaginary-time evolution methods are proposed for
the computation of solitary waves in arbitrary spatial dimensions.
For the first method (with traditional power normalization), the
convergence conditions as well as conditions for optimal
accelerations are derived. In addition, it is shown that for
nodeless solitary waves, this method converges if and only if the
solitary wave is linearly stable. The second method is similar to
the first method except that it uses a novel amplitude
normalization. The performance of these methods is illustrated on
various examples. It is found that while the first method is
competitive with the Petviashvili method, the second method delivers
much better performance than the first method and the Petviashvili
method.


\section{Introduction}

In the study of nonlinear wave equations, solitary waves play an
important role. In certain cases (such as in integrable equations),
solitary waves can be calculated analytically. But in a majority of
other cases, analytical expressions for solitary waves are not
available. Important examples which have arisen recently in the
study of physical systems include optical solitary waves in periodic
media \cite{segev, Kivshar_book, YangMuss03} and nonlinear matter
waves in Bose-Einstein condensates (see, e.g., \cite{BuschA_98,
ChiofaloS_00, CarrC_02}). In such cases, one relies on numerical
techniques to determine the shapes of solitary waves. Several types
of numerical methods have been proposed and used for this purpose,
such as the Newton's iteration method \cite{Keller,Boyd}, the
shooting method \cite{YangPRE02}, the nonlinear Rayleigh-Ritz
iteration method \cite{Panos}, the Petviashvili method \cite{Pet,
Peli_Pet,Ablowitz_Muss,LakobaYang}, the imaginary-time evolution
method \cite{BuschA_98, ChiofaloS_00, CarrC_02, GarciaRipollP_01,
Bao_Du, VS_04}, and the squared-operator iteration methods
\cite{YangLakoba}. The imaginary-time evolution method is attractive
for its simple implementation, insensitivity to the number of
dimensions, and high accuracy (due to its compatibility with the
pseudo-spectral method). In addition, if it converges, it usually
does so faster than the squared-operator iteration methods.

The idea of the imaginary-time evolution method (ITEM) as applied to
linear equations is quite old (see, e.g., \cite{DR_55, Koonin_86}).
In the past decade, this method has also been applied to nonlinear
equations \cite{BuschA_98, ChiofaloS_00, CarrC_02, GarciaRipollP_01,
VS_04}. In this method, one seeks the stationary solution of an
evolution equation (usually, of the parabolic type) by numerically
integrating that equation where time $t$ is replaced by $it$ (hence
the name `imaginary-time'), and normalizing the solution after each
step of time integration to have a fixed $L_2$ norm (called power by
physicists). For linear equations, this method has long been known
(see, e.g., \cite{Koonin_86}) to be equivalent to the problem of
minimizing the energy functional of the physical system under the
constraint that the solution being sought has a given power.
Recently, this same statement was shown to hold for nonlinear
equations as well \cite{VS_04}. In \cite{Bao_Du}, the authors
treated the ITEM as a normalized gradient flow and proved its energy
diminishing property. The ITEM, in its original form, is quite slow.
In addition, it does not always converge to a stationary solution
even if the initial function is quite close to the solution. In an
effort to improve the convergence rate of the ITEM, the authors of
Ref. \cite{GarciaRipollP_01} demonstrated that if the Sobolev
gradients are used in the minimization of the energy functional, the
convergence of the ITEM can be greatly accelerated (an equivalent
possibility was mentioned in passing in \cite{Koonin_86}, although
no related details were provided there.) Alternatively, the authors
of \cite{VS_04} used the steepest descent technique in the
minimization of the energy functional and achieved fast convergence
as well. However, these earlier studies or applications of the ITEM
did not consider the conditions under which the ITEM and its
accelerated versions would converge. So it was not clear when those
methods could be used. In addition, the important practical question
of establishing the conditions for the optimal acceleration of the
ITEM was not considered either.

An important question in the studies of solitary waves is their
linear stability \cite{VK, Jones88, grillakis88, weinstein86,
strauss77}. Most numerical techniques used to find solitary waves,
such as the Newton's iteration method, the shooting method, and the
Petviashvili method, yield no information about the stability of the
solitary wave being obtained \cite{Keller,YangPRE02,Peli_Pet}. A
remarkable fact about the ITEM and its properly accelerated version
(with the usual power normalization), as we will show in this paper,
is that the convergence of this numerical method is directly related
to the linear stability of the corresponding solitary wave, provided
that this wave is nodeless. This means that both existence and
linear stability of these solitary waves can be obtained by this
single numerical procedure.

In this paper, we propose two new accelerated ITEMs for the
computation of solitary waves in general nonlinear wave equations.
Our acceleration technique is to introduce an acceleration operator
to the imaginary-time equation, analogous to the preconditioning
technique for solving systems of linear equations. For the first
method, which uses power normalization, three important theoretical
results are derived. One result is that convergence conditions of
this method are explicitly obtained. This puts the application of
this method on a solid theoretical footing. These convergence
conditions show that in most cases, this method converges when the
underlying solitary wave is nodeless, but there also exist cases
when the method converges to solitary waves with nodes. Another
result is that for nodeless solitary waves, this method converges if
and only if the solitary wave is linearly stable. This connection
between convergence of this method and linear stability of the
underlying solitary wave is a novel property of this numerical
method which has not been seen in other schemes. The third result is
that explicit conditions for optimal acceleration of this method are
obtained. These results provide the optimal practical implementation
of this accelerated ITEM. The performance of this method is
illustrated on various examples, and it is found to be competitive
with the Petviashvili method. The second accelerated ITEM which we
propose is similar to the first method except that it uses a
non-traditional amplitude normalization. We will show through
examples that this second method delivers better performance than
the first method and the Petviashvili method.

This paper is structured as follows. In Sec. 2, we introduce the
original imaginary-time evolution method. In Sec. 3, we propose the
first accelerated imaginary-time evolution method and derive its
convergence conditions. In Sec. 4, we show that for nodeless
solitary waves, the convergence of this first method is directly
linked to the linear stability of the solitary wave. In Sec. 5, we
establish explicit conditions for the optimal acceleration of the
first method. In Sec. 6, we propose the second accelerated
imaginary-time evolution method, which employs the amplitude
normalization. In Sec. 7, we apply both methods as well as the
Petviashvili method to several examples, and show that the second
accelerated imaginary-time evolution method delivers the best
performance, while the first method is comparable to the
Petviashvili method in performance. Sec. 8 concludes the paper. In
the appendix, we attach a matlab code for one of the examples.


\section{Preliminaries on the original imaginary-time evolution method}

The problem we are interested in is the numerical determination of
solitary waves in general scalar nonlinear wave equations in
arbitrary spatial dimensions. To maintain the focus of the
presentation, we first develop the theory for the $N$-dimensional
generalized nonlinear Schr\"odinger equation with an arbitrary
potential. The extension of this theory to more general scalar
nonlinear wave equations will be presented at the end of Sec.
\ref{sec_opt}.

The $N$-dimensional generalized nonlinear Schr\"odinger equation with an arbitrary potential
has the following form:
\begin{equation} \label{NLS}
iU_t+\nabla^2 U+F(|U|^2, {\mathbf x}) U=0,
\end{equation}
where ${\mathbf x}=(x_1, x_2, \dots x_N)$ is a $N$-dimensional
spatial variable,
\begin{equation}
\nabla^2=\frac{\partial^2}{\partial x_1^2}+\frac{\partial^2}{\partial x_2^2}+\dots
+\frac{\partial^2}{\partial x_N^2}
\end{equation}
is the $N$-dimensional Laplacian, and  $F(., .)$ is a real-valued
function. This system is Hamiltonian. Solitary waves of Eq.
(\ref{NLS}) are sought in the form
\begin{equation} \label{solution}
U({\mathbf x}, t)=u({\mathbf x})e^{i\mu t},
\end{equation}
where $u({\mathbf x})$ is a real-valued, localized function, and
$\mu$ is a real parameter called the propagation constant. Then,
from Eqs. (\ref{NLS}) and (\ref{solution}), function $u({\mathbf
x})$ is found to satisfy the equation
\begin{equation} \label{ODE}
L_{00} u=\mu u,
\end{equation}
where
\begin{equation}
L_{00} \equiv \nabla^2+F(u^2, {\mathbf x}).
\end{equation}

Equation (\ref{ODE}) admits solitary waves for a large class of
functions $F(u^2, {\mathbf x})$ \cite{strauss77b,berestycki83}. In
this paper, we always assume that the solitary wave we are trying to
obtain numerically does exist.

In the original imaginary-time evolution method, one numerically
integrates the equation
\begin{equation}  \label{ITM0}
u_t=L_{00}u,
\end{equation}
which is obtained from Eq. (\ref{NLS}) by replacing $t$ with $it$
(hence the name `imaginary-time'), and then normalizes the solution
after each step of time integration to have a fixed power. The power
$P$ of the solitary wave $u({\mathbf x})$ is defined as
\begin{equation}
P(\mu)=\int^\infty_{-\infty} u^2({\mathbf x}; \mu) d{\mathbf x}.
\label{def_P}
\end{equation}
The simplest implementation of numerical time integration is to use the Euler method,
whereby the ITEM scheme is:
\begin{equation} \label{ITEM}
u_{n+1}=\left[\frac{P}{\langle \hat{u}_{n+1}, \hat{u}_{n+1} \rangle}\right]^{\frac{1}{2}}
\hat{u}_{n+1},
\end{equation}
and
\begin{equation} \label{ITEMun}
\hat{u}_{n+1} = u_n+[L_{00}u]_{u=u_n}\Delta t.
\end{equation}
Here $u_n$ is the solution after the $n$th iteration,
\begin{equation} \label{ip}
\langle f, g\rangle = \int^\infty_{-\infty} f({\mathbf x})^*
g({\mathbf x}) d{\mathbf x}
\end{equation}
is the standard inner product in the $N$-dimensional space of
square-integrable functions, and the superscript ``*" represents
complex conjugation.
Note that step (\ref{ITEM}) of the ITEM scheme
guarantees that the power at every iteration is conserved:
\begin{equation}  \label{ununP}
\langle u_n, u_n \rangle =P,   \hspace{1cm} n=1, 2, \dots.
\end{equation}
Thus, if iterations (\ref{ITEM})-(\ref{ITEMun}) converge to
a solitary wave $u({\mathbf x})$, then this $u({\mathbf x})$ must satisfy Eq. (\ref{ODE}),
with its power being $P$ and the propagation constant $\mu$ being equal to
\begin{equation} \label{mu}
\mu=\frac{1}{P}\langle L_{00}u, u \rangle.
\end{equation}

In the following sections, we will use two linear operators, whose
explicit forms are given by the following definition.
\begin{definition} \label{definition1}
Operators $L_0$ and $L_1$ are defined as
\begin{equation}  \label{L0}
L_0 \equiv L_{00}-\mu=\nabla^2+F(u^2, {\mathbf x})-\mu,
\end{equation}
and
\begin{equation}  \label{L}
L_1 \equiv \nabla^2+F(u^2, {\mathbf x})+2u^2F_{u^2}(u^2, {\mathbf
x})-\mu,
\end{equation}
where $F_{u^2}\equiv \partial F/\partial u^2$.
\end{definition}
Under these notations, $L_0u=0$, and $L_1$ is the linearization
operator of $L_0u$ with respect to $u$.


\section{An accelerated imaginary-time evolution method} \label{acceleration}

A major drawback of the original ITEM (\ref{ITEM})-(\ref{ITEMun}) is
that its convergence is quite slow, because the time step $\Delta t$
has to be very small in order for it to converge (see Note 1 below
for explanation). To overcome this difficulty, one idea is to use
implicit time-stepping methods (such as the backward Euler method)
to solve the imaginary-time equation (\ref{ITM0}) (see, e.g.,
\cite{Bao_Du}). Implicit methods allow larger time steps without
causing divergence to the iterations. However, for nonlinear
equations or in high dimensions, implicit schemes are difficult to
implement, and their accuracies are often low (if finite-difference
discretization is used). Another idea of accelerating the ITEM (see
\cite{GarciaRipollP_01}) is to use the Sobolev gradients in the
minimization of the energy functional. This idea is analogous to the
preconditioning technique applied to the imaginary-time equation
(\ref{ITM0}). Below we will extend this idea and propose a new
accelerated ITEM that is explicit but fast-converging.

In our accelerated ITEM, instead of evolving the original
imaginary-time equation (\ref{ITM0}), we evolve the following
``pre-conditioned" imaginary-time equation
\be u_t=M^{-1}\left[ L_{00}u-\mu u \right], \label{t1_08} \ee
where $M$ is a positive-definite and self-adjoint ``preconditioning"
operator. The stationary solution of this equation is still $u(\x)$.
Applying the Euler method to this new equation, the new accelerated
ITEM method (AITEM) we propose is:
\begin{equation}
u_{n+1}=\hat{u}_{n+1} \sqrt{\frac{P}{\langle \hat{u}_{n+1},
\hat{u}_{n+1} \rangle} }\ , \label{t1_09a}
\end{equation}
\be \hat{u}_{n+1}=u_n+ M^{-1}\left(L_{00}u-\mu u \right)_{u=u_n,\
\mu=\mu_n} \Delta t, \label{t1_09} \ee
and
\be \mu_n=\left. \frac{\langle L_{00}u, M^{-1} u \rangle}{\langle u,
M^{-1} u \rangle} \right|_{u=u_n}, \label{t1_10} \ee
where $P$ is the power defined in Eq. (\ref{def_P}), which is
pre-specified. Notice that our updating formula (\ref{t1_10}) for
$\mu_n$ is quite special, different from the usual formula
(\ref{mu}). This special updating formula (\ref{t1_10}) enables us
to derive the convergence properties of the above AITEM, which we
will do in this section. If $M$ is the identity operator, then the
above scheme is closely related to the original ITEM
(\ref{ITEM})-(\ref{ITEMun}), and both have similar (slow)
convergence properties. But if $M$ takes other sensible forms [such
as (\ref{M}) below], convergence of the above AITEM will be much
faster. In this paper, we will call $M$ the acceleration operator.

Before discussing the convergence conditions of the AITEM
(\ref{t1_09a})-(\ref{t1_10}), we first present an example to
demonstrate the drastic improvement in convergence of the AITEM
(\ref{t1_09a})-(\ref{t1_10}) over that of the ITEM
(\ref{ITEM})-(\ref{ITEMun}).

\textbf{Example 1} \ The nonlinear Schr\"odinger equation in one
spatial dimension,
\begin{equation} \label{NLS0}
u_{xx}+u^3=\mu\, u
\end{equation}
admits the solitary wave
\begin{equation}
u(x)=\sqrt{2}\: \mbox{sech} x \label{sol_NLS}
\end{equation}
with $P=4$ and $\mu=1$. To apply the AITEM for this solution, we
take $M=1-\partial_{xx}$, $\Delta t=1.5$, and the Gaussian initial
condition $u_0(x)=e^{-x^2}$. The $x$-interval is taken to be $[-15,
15]$, discretized by 128 grid points, and the discrete Fourier
transform is used to calculate $u_{xx}$ and to invert operator $M$.
For the ITEM, the scheme parameters are the same, except that $\D
t=0.01$ (when $\Delta t > 0.011$, the ITEM diverges). Defining the
error as the $L_2$ norm of the solution difference between
successive iterations, i.e. $[\int (u_n-u_{n-1})^2 dx]^{1/2}$, we
find that for the error to drop below $10^{-10}$, the AITEM and ITEM
take, respectively, 33 and 2160 iterations. Thus, the AITEM for this
case is about two orders of magnitude faster than the original ITEM.
The main reason for this drastic improvement of convergence is that
for the AITEM, the time step $\Delta t$ can be taken much larger
(which is 1.5 above) without causing divergence. This is made
possible by the introduction of the acceleration operator $M$.

Now we derive the convergence conditions for the AITEM
(\ref{t1_09a})-(\ref{t1_10}). To do so, we introduce the following
assumption on the kernel of $L_1$ which holds in generic cases (in
rare cases, this assumption may break down, see Fig. 7 in
\cite{YangChen_defectsoliton}).

\begin{assumption} \label{assumption1}
If function $F$ in Eq. (\ref{ODE}) does not depend explicitly on
certain spatial coordinates $\{x_{j_1}, x_{j_2}, ..., x_{j_k}\}$ ($1
\le j_1, j_2, ..., j_k \le N$), we assume that the only
eigenfunctions in the kernel of $L_1$ that are orthogonal to
$u({\mathbf x})$ are the $k$ translational-invariance modes
$\partial u/\partial x_{j_s}({\mathbf x}), 1\le s \le k$.
\end{assumption}

We also introduce a notation: for any operator $L$, we denote the
number of its positive eigenvalues as $p(L)$.

Under the above assumption and notation, we have the following
theorem on the convergence of the AITEM
(\ref{t1_09a})-(\ref{t1_10}).

\begin{theorem}
\label{theorem1} Let Assumption \ref{assumption1} be valid. Define
$\D t_{max}=-2/\Lambda_{min}$, where $\Lambda_{min}$ is the minimum
(negative) eigenvalue of operator $\cL$ in Eq. (\ref{t1_15}). Then
if $\Delta t > \Delta t_{max}$, the AITEM
(\ref{t1_09a})-(\ref{t1_10}) diverges. If $\Delta t < \Delta
t_{max}$, the following convergence statements on this AITEM hold.
\begin{enumerate}
\item
If $p(L_1)=0$ and $P'(\mu) \ne 0$,
 then the AITEM converges.
\item
Suppose $p(L_1)=1$, then $p(M^{-1}L_1)=1$. Denote the eigenfunction
of this single positive eigenvalue of $M^{-1}L_1$ as
 $\psi({\mathbf x})$. Then if $\langle \psi, u \rangle \ne 0$,
 the AITEM converges for $P'(\mu)>0$ and diverges for $P'(\mu)<0$.
 If, however, $\langle \psi, u \rangle = 0$, the AITEM diverges.
\item If $p(L_1)> 1$, the AITEM diverges.
\end{enumerate}
\end{theorem}

\noi {\bf Proof} \ To analyze the convergence properties of the
AITEM (\ref{t1_09a})-(\ref{t1_10}), we use the linearization
technique. Let
\begin{equation}  \label{linearize}
u_{n}=u+\tilde{u}_n, \qquad | \tilde{u}_n | \ll |u|,
\end{equation}
where $\tilde{u}_n(\x)$ is the error. When this equation is
substituted into the power-normalization step (\ref{t1_09a}) and
only terms of $O(\tilde{u}_n)$ retained, one obtains that the error
is orthogonal to $u({\mathbf x})$:
\begin{equation}  \label{un_u}
\langle \tilde{u}_n,  u \rangle=0, \qquad \mbox{for all $n$}.
\end{equation}
Substituting Eq. (\ref{linearize}) into (\ref{t1_09})-(\ref{t1_10})
and linearizing, we find that the error satisfies the following
iteration equation
\be \ut_{n+1}=(1 + \D t \,\cL) \ut_n , \label{t1_14} \ee
where operator $\cL$ is
\be \cL\Psi = M^{-1} \left( L_1 \Psi - \frac{ \langle L_1\Psi,
M^{-1}u \rangle }{\langle u, M^{-1} u \rangle} u\,\right) .
\label{t1_15} \ee
Convergence of the AITEM depends on the eigenvalues of operator
$\cL$. In view of the orthogonality constraint (\ref{un_u}), the
eigenvalue problem for ${\cal L}$ that we need to consider is
\begin{equation} \label{operator}
{\cal L}\Psi =\Lambda \Psi, \hspace{0.5cm} \Psi \in S,
\end{equation}
where
\begin{equation} \label{S} S=\{\Psi({\mathbf x}): \langle
\Psi, u \rangle =0 \}.
\end{equation}
Note that when $\Lambda \ne 0$, by taking the inner product between
equation ${\cal L}\Psi =\Lambda \Psi$ and $u$, one easily gets
$\langle \Psi, u\rangle=0$, i.e. $\Psi \in S$. Thus $\Psi\in S$ in
Eq. (\ref{operator}) constitutes a constraint only for zero
eigenvalues of ${\cal L}$. As we will show below, all eigenvalues
$\Lambda$ of $\cL$ are real, and all eigenfunctions of $\cL$ form a
complete set in $S$. Thus, in view of Assumption 1, the necessary
and sufficient conditions for the AITEM to converge are that (i) for
all non-zero eigenvalues $\Lambda$ of $\cL$,
\begin{equation} \label{ineq}
-1 < 1+\Lambda \Delta t <1;
\end{equation}
(ii) for the zero eigenvalue of Eq. (\ref{operator}) (if exists),
its eigenfunctions must be translational-invariance eigenmodes
$u_{x_j}$ (which lead to a spatially translated solitary wave of Eq.
(\ref{ODE}) and do not affect the convergence of iterations). If
$\Delta t > \Delta t_{max}$ with $\Delta t_{max}$ given in Theorem
\ref{theorem1}, then the left inequality in (\ref{ineq}) is not met,
thus the AITEM diverges. If $\Delta t < \Delta t_{max}$, the left
inequality in (\ref{ineq}) is satisfied, hence we only need to
consider the right inequality in (\ref{ineq}), and the second
condition (ii).

We consider the condition (ii) first. Suppose Eq. (\ref{operator})
has a zero eigenvalue with eigenfunction $\Psi(\x)$, i.e.,
\begin{equation}
L_1 \Psi - \frac{ \langle L_1\Psi, M^{-1}u \rangle }{\langle u,
M^{-1} u \rangle} u=0, \quad \Psi \in S. \label{t03_02}
\end{equation}
Differentiating Eq. (\ref{ODE}) with respect to parameter $\mu$, we
find that
 \begin{equation} \label{L1umu}
  L_1 u_{\mu}=u.
  \end{equation}
Hence the solution $\Psi$ of Eq. (\ref{t03_02}) is $\alpha u_\mu$,
where $\alpha$ is a constant, plus functions in the kernel of $L_1$.
In view of Assumption 1 on the kernel of $L_1$, we see that $\langle
\Psi, u\rangle = \alpha \langle u_{\mu}, u \rangle=\alpha P'(\mu)
/2$. Thus if $P'(\mu)\neq 0$, then in order for $\Psi\in S$,
$\alpha$ must be zero, hence $\Psi$ is in the kernel of $L_1$.
According to Assumption 1, $\Psi$ then must be a
translational-invariance eigenmode which does not affect the
convergence of iterations.

Next we consider the right inequality in (\ref{ineq}), which is
simply $\Lambda <0$. We will analyze when this condition is met. To
facilitate the analysis, we introduce the following new variables
and operators
\be \hat{\Psi}=M^{1/2}\Psi , \quad \hat{u}= M^{-1/2} u, \quad
\hat{L}_{1}= M^{-1/2} L_1 M^{-1/2}. \label{t1_17} \ee
Then the eigenvalue problem (\ref{operator}) becomes the following
equivalent one with eigenvalues unchanged:
\begin{equation} \label{cLhat}
\hat{\cal L}\hat{\Psi}=\Lambda \hat{\Psi}, \hspace{0.5cm} \hat{\Psi}
\in \hat{S},
\end{equation}
where
\begin{equation} \label{cLhat2}
\hat{\cal L}\hat{\Psi} =\hat{L}_1 \, \hat{\Psi} - \hat{H} \hat{u},
\hspace{0.5cm} \hat{H}=\frac{\langle \hat{L}_{1} \,\hat{\Psi},
\hat{u} \rangle}{\langle \hat{u}, \hat{u} \rangle},
\end{equation}
and
\begin{equation}\label{S_hat}
\hat{S}=\{\hat{\Psi}({\mathbf x}): \langle \hat{\Psi}, \hat{u}
\rangle =0 \}.
\end{equation}
It is easy to see that operator $\hat{\cL}$ is self-adjoint in the
set $\hat{S}$ due to $L_1$ and $M$ being self-adjoint and $M$ being
positive definite. Thus all eigenvalues $\Lambda$ of $\hat{\cL}$ are
real, and all eigenfunctions $\hat{\Psi}$ of $\hat{\cL}$ form a
complete set in $\hat{S}$. As a result, all eigenvalues $\Lambda$ of
$\cL$ are real, and all eigenfunctions of $\cL$ form a complete set
in $S$. In addition, since $\hat{L}_1$ is similar to $M^{-1}L_1$ and
in view of the {\em Sylvester inertia law} (see, e.g., Theorems
4.5.8 and 7.6.3 in \cite{HornJohnson91}),
$p(\hat{L}_1)=p(M^{-1}L_1)=p(L_1)$.

The eigenvalue problem (\ref{cLhat}) is equivalent to the one
arising in the linear stability analysis of nodeless solitary waves
(see, e.g., \cite{Kivshar_book, VK} and, in particular, Eq. (2.3.9)
in \cite{Kivshar_book}), and has been well studied. To determine the
sign of eigenvalue $\Lambda$, we expand $\hat{\Psi}$ into the
complete set of eigenfunctions of the self-adjoint operator
$\hat{L}_1$ as
\begin{equation} \label{Psihat}
\hat{\Psi}({\mathbf x})=\sum_{k} b_k \hat{\psi}_k({\mathbf x}) +
\int_I b(\lambda)\hat{\psi}({\mathbf x}; \lambda)d\lambda,
\end{equation}
where $\hat{\psi}_k({\mathbf x})$ and $\hat{\psi}({\mathbf x};
\lambda)$ are the properly normalized discrete and continuous
eigenfunctions of $\hat{L}_1$ with eigenvalues $\lambda_k$ and
$\lambda$. The coefficients in Eq. (\ref{Psihat}) are given by
$b_k=\langle \hat{\psi}_k, \hat{\Psi} \rangle$ and $b=\langle
\hat{\psi}, \hat{\Psi} \rangle$. Function $\hat{u}$ can be expanded
in a similar way with coefficients $c_k=\langle \hat{\psi}_k,
\hat{u} \rangle$ and $c(\lambda)=\langle \hat{\psi}, \hat{u}
\rangle$. When $\hat{H}=0$, the analysis is trivial and the
convergence conditions in Theorem \ref{theorem1} can be easily
obtained. Thus we only consider the $\hat{H} \ne 0$ case below. In
this case, substituting expansions of $\hat{u}$ and $\hat{\Psi}$
into Eq. (\ref{cLhat}), one finds that $b_k=\hat{H} c_k
/(\lambda_k-\Lambda), b(\lambda)=\hat{H}
c(\lambda)/(\lambda-\Lambda)$. Substituting these relations into the
orthogonality condition $\langle \hat{\Psi}, \hat{u} \rangle =0$, we
get
\begin{equation} \label{Q}
Q(\Lambda) \equiv \sum_k \frac{|c_k|^2}{\lambda_k-\Lambda} + \int_I
\frac{|c(\lambda)|^2}{\lambda-\Lambda}d\lambda =0.
\end{equation}
If $p(L_1)=p(\hat{L}_1)=0$, i.e. all eigenvalues of $\hat{L}_1$ are
negative, then $Q(\Lambda)$ does not change sign when $\Lambda> 0$,
hence equation (\ref{Q}) has no positive roots, and the right
inequality in (\ref{ineq}) is met. We have shown above that when
$P'(\mu)\ne 0$, condition (ii) is met also, thus the AITEM
converges. If $p(L_1)=p(\hat{L}_1)> 1$, denote $\hat{L}_1$'s two
positive eigenvalues as $\lambda_1$ and $\lambda_2$. If the
corresponding expansion coefficients $c_1$ and $c_2$ in $\hat{u}$
are such that $c_1 c_2\ne 0$, then since $Q(\Lambda)$ is continuous
and monotonic in the interval $(\lambda_1, \lambda_2)$, and it
approaches $-\infty$ and $\infty$ at the two ends of this interval,
$Q(\Lambda)$ obviously has a positive root between $\lambda_1$ and
$\lambda_2$, hence the right inequality in (\ref{ineq}) is not met,
and the AITEM diverges. If $c_k = 0$ ($k=1$ or 2), then
$\Lambda=\lambda_k> 0$ is an eigenvalue of Eq. (\ref{cLhat}) with
$\hat{\Psi}=\hat{\psi}_k$, hence the AITEM also diverges.

Now we consider the case $p(L_1)=p(\hat{L}_1)=1$. Denote this single
positive eigenvalue of $\hat{L}_1$ as $\lambda_1$, and its
eigenfunction as $\hat{\psi}$. If $\langle \hat{\psi},
\hat{u}\rangle=0$, then $\Lambda=\lambda_1> 0$ is an eigenvalue of
Eq. (\ref{cLhat}) with $\hat{\Psi}=\hat{\psi}$, hence the AITEM
diverges. Below we consider the $\langle \hat{\psi}, \hat{u}\rangle
\ne 0$ case. Clearly $Q(\Lambda)$ cannot have zeros when $\Lambda >
\lambda_1$. Whether $Q(\Lambda)$ has a positive zero in $(0,
\lambda_1)$ depends on the sign of $Q(0)/\hat{H}$: if
$Q(0)/\hat{H}<0$, then $Q(\lambda)$ has a positive zero, and vise
versa. From Eqs. (\ref{cLhat}) and (\ref{cLhat2}), we see that
$Q(0)/\hat{H}=\langle \hat{L}_1^{-1}\hat{u}, \hat{u}\rangle$. In
view of Eq. (\ref{t1_17}), it follows that $Q(0)/\hat{H}=\langle
L_1^{-1}u, u\rangle$. Then from Eq. (\ref{L1umu}), we get
$Q(0)/\hat{H}=P'(\mu)/2$. Consequently, if $P'(\mu)>0$, the right
inequality in (\ref{ineq}) is met, and the AITEM converges; if
$P'(\mu)<0$, the AITEM diverges. Lastly, we notice from Eq.
(\ref{t1_17}) that when $\hat{\psi}$ is an eigenfunction of
$\hat{L}_1$, $\psi=M^{-1/2}\hat{\psi}$ is an eigenfunction of
$M^{-1}L_1$ at the same eigenvalue, thus $\langle \hat{\psi},
\hat{u}\rangle=\langle \psi, u\rangle$. This completes the proof of
Theorem \ref{theorem1}. $\Box$

\textbf{Note 1} \ If $M$ is taken as the identity operator, then our
AITEM (\ref{t1_09a})-(\ref{t1_10}) becomes similar to the original
ITEM (\ref{ITEM})-(\ref{ITEMun}). In this case, the smallest
eigenvalue of $\cL$ is $\Lambda_{min}=-\infty$, which leads to
$\Delta t_{max}=0$. However, in any computer implementation of this
method, space is discretized. For the discretized operator of $\cL$,
$\Lambda_{min}$ is finite, not $-\infty$, thus $\Delta t_{max}>0$,
hence this method (as well as the original ITEM) can still be used
(see Example 1). But for any accurate spatial discretizations,
$\Lambda_{min}$ is large negative, hence $\Delta t_{max}$ is very
small, which results in the slow convergence of this method and the
original ITEM. For a more sensible choice of $M$ such as (\ref{M})
below, the smallest eigenvalues of $\cL$ and its discretized version
are both $O(1)$ and almost identical. This makes $\Delta t_{max}$
much larger, hence the AITEM (\ref{t1_09a})-(\ref{t1_10}) converges
much faster (see Example 1).

Theorem \ref{theorem1} is one of the main results of this article.
It puts the application of the AITEM (\ref{t1_09a})-(\ref{t1_10}) on
a firm theoretical basis. A corollary of this theorem can be readily
established below.
\begin{corollary} \label{corollary2}
Consider Eq. (\ref{ODE}) with $\lim_{\bf |x| \To \infty} F(0, {\bf
x})=0$. Under Assumption 1 and restriction $\Delta t < \Delta
t_{max}$, where $\Delta t_{max}$ is given in Theorem \ref{theorem1},
the AITEM (\ref{t1_09a})-(\ref{t1_10}) diverges if the solitary wave
$u(\x)$ has nodes (where $u({\bf x})=0$) and $F_{u^2}(u^2, \x) \ge
0$ for all $\x$; if $u(\x)$ is nodeless and $F_{u^2}(u^2, \x) \le 0$
for all $\x$, then the AITEM converges.
\end{corollary}

To prove this corollary, we need the following lemma.
\begin{lemma} \label{lemma1}
Consider the $N$-dimensional, linear Schr\"odinger eigenvalue
problem
\begin{equation} \label{ND_schroedinger}
\nabla^2\psi-V({\mathbf x})\psi = \lambda \psi,
\end{equation}
where $V({\mathbf x}) \to 0$ as $|{\mathbf x}| \to \infty$. Let its
discrete eigenvalues be arranged in the decreasing order: $\lambda_1
> \lambda_2 \ge \dots \ge \lambda_m >0$, with the continuous
spectrum being at $\lambda < 0$. Let $V_1(\x)$ and $V_2(\x)$ be two
potentials such that $V_2({\mathbf x}) \le V_1({\mathbf x})$ for all
${\mathbf x}$, and $V_2({\mathbf x}) \not\equiv V_1({\mathbf x})$.
Then the discrete eigenvalues $\{\lambda_k^{(2)}\}$ of Eq.
(\ref{ND_schroedinger}) with potential $V_2$ are larger than the
corresponding eigenvalues $\{\lambda_k^{(1)}\}$ of the same equation
with potential $V_1$, i.e., $\lambda_k^{(2)}
> \lambda_k^{(1)}, k=1, 2, \dots$.
\end{lemma}

This lemma says that deepening the potential (making $V(\x)$ larger)
shifts the eigenvalues of Eq. (\ref{ND_schroedinger}) downward.

\textbf{Proof of Lemma \ref{lemma1}:} \ Define the potential
function
\begin{equation} \label{Valpha}
V({\mathbf x}; \alpha)=V_1({\mathbf x})+\alpha[V_2({\mathbf
x})-V_1({\mathbf x})],
\end{equation}
where $\alpha$ is a real parameter. As $\alpha$ increases from 0 to
1, $V({\mathbf x}; \alpha)$ changes from $V_1({\mathbf x})$ to
$V_2({\mathbf x})$. We now analyze how a discrete eigenvalue
$\lambda$ of Eq. (\ref{ND_schroedinger}) changes as $\alpha$
continuously varies. For this purpose, we differentiate Eq.
(\ref{ND_schroedinger}) with respect to $\alpha$, and obtain
\begin{equation}
\left(\nabla^2-V-\lambda\right)\frac{\partial \psi}{\partial
\alpha}= \frac{d \lambda}{d \alpha} \psi + (V_2-V_1)\psi.
\end{equation}
In order for this equation to have a localized solution $\partial
\psi / \partial \alpha$, its right-hand side must be orthogonal to
the homogeneous solution $\psi({\mathbf x}; \alpha)$. This yields
the relation
\begin{equation}
\frac{d \lambda}{d \alpha}=\frac{\langle (V_1-V_2)\psi, \psi
\rangle}{\langle \psi, \psi \rangle}.
\end{equation}
According to our assumption, $V_2({\mathbf x}) \le V_1({\mathbf x})$
for all ${\mathbf x}$, and $V_2({\mathbf x}) \ne V_1({\mathbf x})$;
thus $d \lambda /d \alpha>0$. This means that $\lambda_k^{(2)}
> \lambda_k^{(1)}, k=1, 2, \dots$. Hence Lemma \ref{lemma1} is
proved. $\Box$

With this lemma, we now prove Corollary \ref{corollary2}.

{\bf Proof of Corollary \ref{corollary2}}: \ First, we recall the
definitions of the two operators $L_0$ and $L_1$ in Eqs. (\ref{L0})
and (\ref{L}), and the fact of $L_0 u(\x) =0$, i.e., $L_0$ has a
zero eigenvalue $\lambda_a=0$ with eigenfunction $u(\x)$. We will
also use a well known result about linear Schr\"odinger operators,
which says that the largest eigenvalue of those operators is simple
and the corresponding eigenfunction is nodeless \cite{Struwe00}. We
now consider two possibilities in regards to the nodes of $u(\x)$.

(1) Suppose $u(\x)$ has at least one node. Then, by the
aforementioned property of linear Schr\"odinger operators, $L_0$
must also have at least one positive eigenvalue $\lambda_b>0$ whose
eigenfunction is nodeless. Comparing the two Schr\"odinger operators
$L_0$ and $L_1$, we see that the difference in their potentials is
$2u^2 F_{u^2}(u^2, \x)$. If $F_{u^2}(u^2, \x) \ge 0$ and
$F_{u^2}(u^2, \x)\not\equiv 0$, then it is seen from Lemma
\ref{lemma1} that operator $L_1$ must have at least two positive
eigenvalues corresponding to the eigenvalues $\lambda_a$ and
$\lambda_b$ of $L_0$. Hence according to Theorem \ref{theorem1}, the
AITEM diverges.

(2) If $u(\x)$ is nodeless, then zero is the largest eigenvalue of
$L_0$. When $F_{u^2}(u^2, \x) \le 0$ and $F_{u^2}(u^2, \x)\not\equiv
0$, then by Lemma \ref{lemma1}, the eigenvalues of $L_1$ are all
negative. Thus according to Theorem \ref{theorem1}, the AITEM
converges. This completes the proof of Corollary \ref{corollary2}.
$\Box$

Corollary \ref{corollary2} can be readily used on certain equations,
and two examples are shown below.

\begin{figure}[h]
\begin{center}
\parbox{12cm}{\postscript{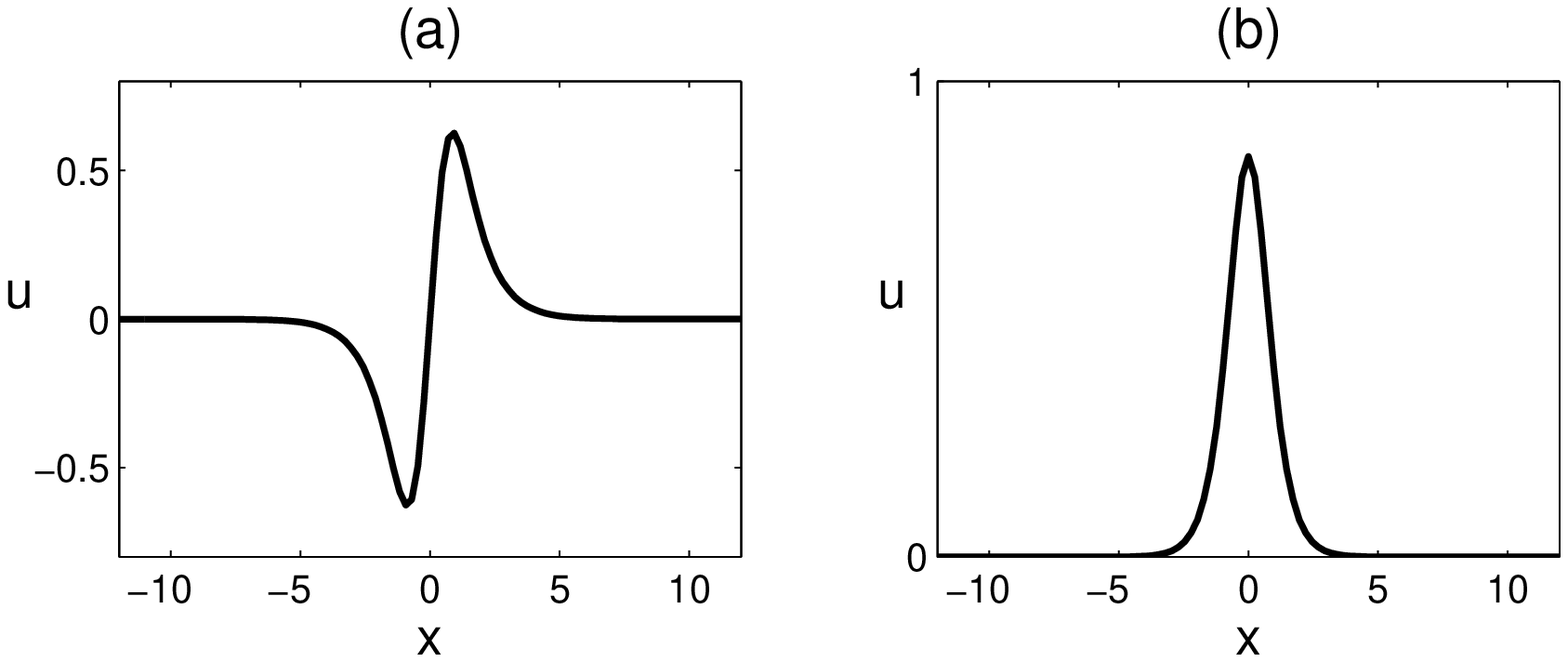}{1.0}}

\caption{(a) A solitary wave with nodes in the focusing nonlinear
Schr\"odinger equation (\ref{1Du}) ($\mu=1.2689, P=1$), for which
the AITEM (\ref{t1_09a})-(\ref{t1_10}) diverges for any $\Delta t$;
(b) A nodeless solitary wave in the defocusing nonlinear
Schr\"odinger equation (\ref{1Du2}) ($\mu=3.5069, P=1$), for which
the AITEM (\ref{t1_09a})-(\ref{t1_10}) converges if the stepsize
condition $\Delta t< \Delta t_{max}$ in Theorem 1 is met.
\label{fig1}}
\end{center}
\end{figure}

1. Consider the focusing nonlinear Schr\"odinger equation with a
single-well potential:
\begin{equation}\label{1Du}
u_{xx}+6\mbox{sech}^2x \; u+u^3=\mu u.
\end{equation}
This equation admits a family of solitary waves with nodes [see an
example in Fig. \ref{fig1}(a)]. Here $F_{u^2}(u^2, \x)=1>0$, thus
according to Corollary \ref{corollary2}, the AITEM diverges for
these solutions, which we have confirmed numerically.

2. Consider the defocusing nonlinear Schr\"odinger equation with a
single-well potential:
\begin{equation}\label{1Du2}
u_{xx}+6\mbox{sech}^2x \; u-u^3=\mu u.
\end{equation}
This equation admits a family of nodeless solitary waves [see an
example in Fig. \ref{fig1}(b)]. Here $F_{u^2}(u^2, \x)=-1<0$, thus
according to Corollary \ref{corollary2}, the AITEM converges for
these solutions under restriction $\Delta t< \Delta t_{max}$ (even
though it can be verified that $P'(\mu)<0$ here).

Corollary \ref{corollary2} suggests that the AITEM tends to diverge
for solitary waves with nodes (such solutions are often called
``excited states"). Indeed, this if often the case. However, there
are examples where the AITEM {\it converges} for solitary waves with
nodes. This could occur if all the eigenvalues of $L_1$ are
negative. For a solitary wave $u({\mathbf x})$ with nodes, since
$L_0 u=0$, $L_0$ must also have positive eigenvalues
\cite{Struwe00}. But if $F_{u^2}(u^2, {\mathbf x}) \le 0$ for all
${\mathbf x}$, then the eigenvalues of $L_1$ are lower than those of
$L_0$ according to Lemma \ref{lemma1}. If the potential difference
$|2u^2F_{u^2}(u^2, {\mathbf x})|$ between $L_0$ and $L_1$ is large
enough, it is possible for all the eigenvalues of $L_1$ to be pushed
below zero, resulting in the convergence of the AITEM. One such
example is given below.

\begin{figure}[h]
\begin{center}
\parbox{12cm}{\postscript{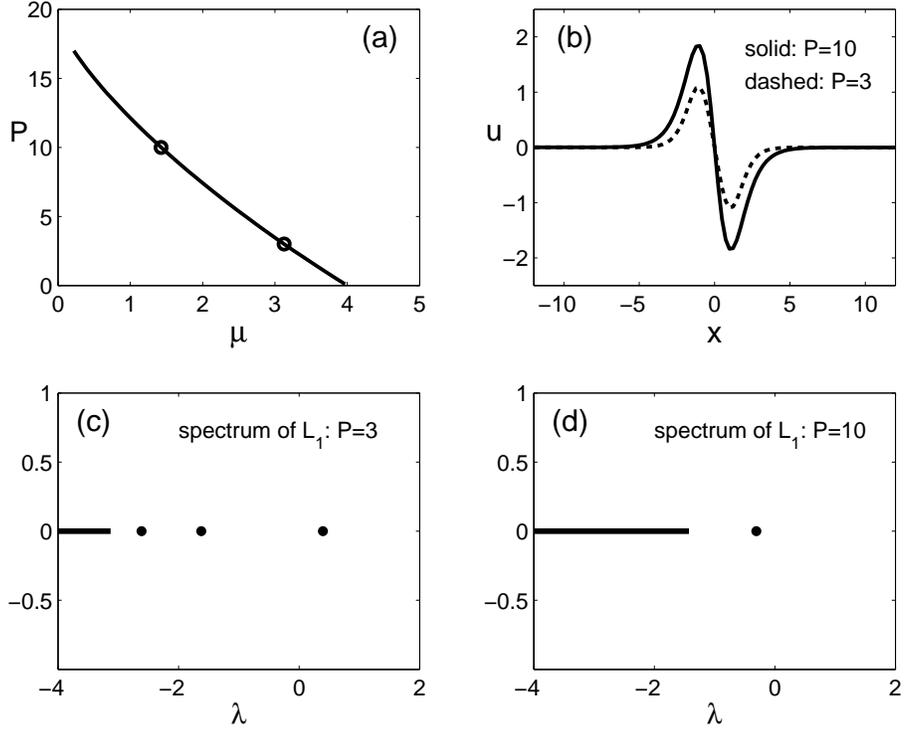}{1.0}}

\caption{Solitary waves with nodes in the defocusing nonlinear
Schr\"odinger equation (\ref{1Du_double}) and their $L_1$ spectra.
(a) The power diagram; (b) two solitary waves with powers $P=3$ and
10; (c, d) spectra of $L_1$ for these two waves. The AITEM
(\ref{t1_09a})-(\ref{t1_10}) converges for the one with $P=10$ when
the stepsize restriction in Theorem \ref{theorem1} is met.
\label{nodes_soliton}}
\end{center}
\end{figure}

{\bf Example 2} \ Consider the defocusing nonlinear Schr\"odinger
equation with a double-well potential:
\begin{equation}\label{1Du_double}
u_{xx}+6\left[\mbox{sech}^2(x+1)+\mbox{sech}^2(x-1)\right]u-u^3=\mu u.
\end{equation}
For this equation, $F_{u^2}(u^2, x)=-1 < 0$. This equation admits a
family of single-node solitary waves, whose power curve is displayed
in Fig. \ref{nodes_soliton}(a). It is seen that $P'(\mu) < 0$ for
the entire family. Two representative solutions with powers $P=3$
and $P=10$ are displayed in Fig. \ref{nodes_soliton}(b). The spectra
of operator $L_1$ for these two waves are plotted in Fig.
\ref{nodes_soliton}(c, d), respectively. At the lower power $P=3$,
$p(L_1)=1$. However, at the higher power $P=10$, $|2u^2F_{u^2}(u^2,
{\mathbf x})|=2u^2$ is large enough such that $p(L_1)=0$, hence
according to Theorem \ref{theorem1}, the AITEM converges.


\section{Connection between convergence and linear stability} \label{stability}

The convergence theorem \ref{theorem1} strongly resembles the linear
stability conditions of nodeless solitary waves in the generalized
NLS equations (\ref{NLS}) \cite{Kivshar_book, VK, grillakis88,
weinstein86, strauss89}. For such solitary waves, the following
stability conditions have been established \cite{Kivshar_book}:

\emph{For a nodeless solitary wave in Eq. (\ref{NLS}), (i) if
$p(L_1)<0$, the wave is linearly stable; (ii) if $p(L_1)=1$, the
wave is linearly stable if $P'(\mu)>0$ and unstable if $P'(\mu)<0$;
(iii) if $p(L_1)>1$, the wave is linearly unstable.}

This stability result is almost identical to our convergence theorem
\ref{theorem1}, indicating that the AITEM
(\ref{t1_09a})-(\ref{t1_10}) usually converges to a nodeless
solitary wave if and only if this wave is linearly stable. The only
notable difference between the above stability and convergence
results is in the case of $p(L_1)=1$, where the convergence theorem
has a condition on $\langle \psi, u \rangle$ which is absent in the
stability theorem. However, for nodeless solitary waves $u(\x)$ with
$p(L_1)=1$, condition $\langle \psi, u \rangle \ne 0$ in Theorem
\ref{theorem1} is met in generic cases, thus the stability and
convergence results are the same for this case as well. For the
following two choices of the acceleration operator $M$, we can
actually show that condition $\langle \psi, u \rangle \ne 0$ is
strictly satisfied.

The first choice is when $M$ is the identity operator, where the
AITEM becomes similar to the original ITEM
(\ref{ITEM})-(\ref{ITEMun}). In this case, if $p(L_1)=1$, then
$\psi(\x)$ is the eigenfunction of the largest eigenvalue of the
Schr\"odinger operator $L_1$, which is known to be nodeless
\cite{Struwe00}. Since $u(\x)$ is nodeless as well, $\langle \psi, u
\rangle$ thus is non-zero.

The second choice is a very practical one, $M=\mu-\nabla^2$, which
will be shown to yield optimal acceleration for a large class of
equations in the next section. For this $M$, the eigenvalue equation
$M^{-1}L_1\psi=\lambda \psi$ can be rewritten as the following
Schr\"odinger equation
\begin{equation} \label{sch2}
\nabla^2\psi-\mu\psi+\frac{\cal V(\x)}{1+\lambda}\psi=0,
\end{equation}
where
\begin{equation} \label{Vdef}
{\cal V}({\mathbf x})= F(u^2, {\mathbf x})+2u^2F_{u^2}(u^2, {\mathbf
x}).
\end{equation}
When $p(L_1)=p(M^{-1}L_1)=1$ and $\lambda$ is the single positive
eigenvalue of $M^{-1}L_1$ [i.e. Eq. (\ref{sch2})], it is easy to
show using the spectral properties of Schr\"odinger operators that
the corresponding eigenfunction $\psi(\x)$ is nodeless (a similar
fact for the 1D case can be found in \cite{Ince}). Since $u(\x)$ is
also nodeless, $\langle \psi, u \rangle\ne 0$.

It is remarkable that the AITEM  (\ref{t1_09a})-(\ref{t1_10}) can
not only produce solitary waves, but also determine their linear
stability properties. This is like ``killing two birds with one
stone". To the authors' knowledge, there are no other numerical
methods for solitary waves which possess this same property. Note,
however, that this property holds only for nodeless solitary waves.
For solitary waves with nodes, this connection between convergence
of the AITEM and linear stability of the solitary wave can break
down.


\section{Optimal acceleration of the imaginary-time evolution method} \label{sec_opt}

In the AITEM (\ref{t1_09a})-(\ref{t1_10}) for Eq. (\ref{ODE}), a
practical choice of the acceleration operator $M$ is
\begin{equation} \label{M}
M=c-\nabla^2, \quad c >0.
\end{equation}
The reason for this choice is two fold: first, $M^{-1}$ is very easy
to compute by the fast Fourier transform; second, all eigenvalues of
$\cL$ are $O(1)$, which makes $\Delta t_{max}=O(1)$ as well. For
this $M$, an important question then is: at what value of $c$ does
the AITEM converge the fastest? We will answer this question in this
section. Note that the use of a seemingly more general form of the
acceleration operator $M=c_1-c_2\nabla^2$ is equivalent to (\ref{M})
by a rescaling of the time step $\Delta t$ in Eq. (\ref{t1_09}), and
hence does not warrant consideration.

First, we quantify the convergence rate of the AITEM. From Eq.
(\ref{t1_14}), we see that the error $\tilde{u}_n$ evolves in
proportion to $R^n$, where
\begin{equation}
R(c, \Delta t) \, = \, \max \left\{ \left| 1 + \Lambda_{min} \D t
\right|, \,
 \left| 1 + \Lambda_{max}\D t\right| \right\},
\label{RAITEM}
\end{equation}
and $\Lambda_{min}(c)$, $\Lambda_{max}(c)$ are the smallest and
largest non-zero eigenvalues of Eq. (\ref{operator}). If $R<1$, the
AITEM converges, and vise versa. In this section, we assume that the
AITEM converges (under the stepsize restriction in Theorem
\ref{theorem1}). This means that both $\Lambda_{min}(c)$ and
$\Lambda_{max}(c)$ are negative. The parameter $R$ characterizes the
rate of convergence and will be called the convergence factor.
Smaller $R$ leads to faster convergence. For fixed $c$, it is seen
from Eq. (\ref{RAITEM}) that the smallest $R$ is reached at
\begin{equation}
\D t = \D t_*(c) \equiv -\frac{2}{\Lambda_{min}+\Lambda_{max}},
\label{DT*AITEM}
\end{equation}
whence
\begin{equation}
R_*(c) \equiv R(c, \Delta t_*) = \frac{\Lambda_{min}-\Lambda_{max}}
{\Lambda_{min}+\Lambda_{max}}\,. \label{R*AITEM}
\end{equation}
The value of $c$ which makes $R_*(c)$ minimal gives optimal
acceleration of the AITEM and will be denoted as $c_{opt}$. The
determination of $c_{opt}$ is the focus of this section.

\begin{figure}[h]
\begin{center}
\parbox{12cm}{\postscript{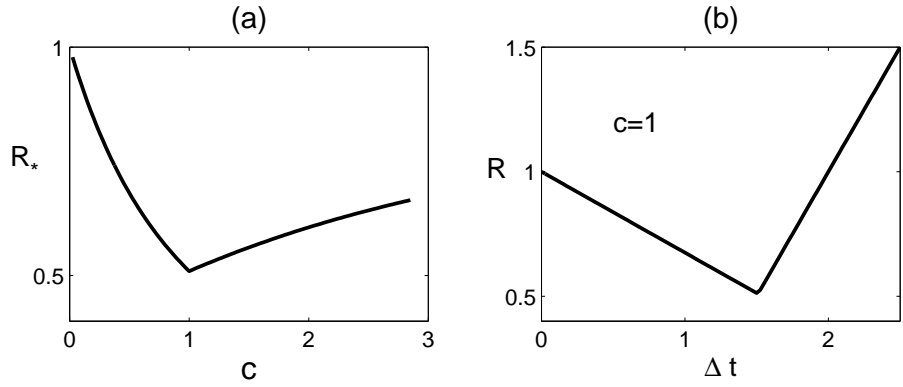}{1.0}}

\caption{Convergence rates of the AITEM (\ref{t1_09a})-(\ref{t1_10})
for the NLS equation (\ref{NLS0}) with $\mu=1$: (a) graph of the
convergence factor $R_*(c)$; (b) graph of $R(\Delta t; c=1)$ versus
$\Delta t$. \label{fig_example}}
\end{center}
\end{figure}

Before analytically determining $c_{opt}$, we first present a
numerical example. Consider the NLS equation (\ref{NLS0}) with
$\mu=1$ again. For each $c$ value in Eq. (\ref{M}), we have
numerically obtained $\Lambda_{min}(c)$ and $\Lambda_{max}(c)$ of
operator $\cL$ by discretizing Eq. (\ref{operator}) and turning it
into a matrix eigenvalue problem. The resulting $R_*(c)$ function is
then obtained from Eq. (\ref{R*AITEM}) and plotted in Fig.
\ref{fig_example}(a). We see that the minimum of $R_*(c)$ occurs at
$c=1$, thus $c_{opt}=1$. At this $c$ value, dependence of the
convergence factor $R$ on the timestep $\Delta t$ can be calculated
from Eq. (\ref{RAITEM}) and is displayed in Fig.
\ref{fig_example}(b). We see that when $\Delta t>2$, $R>1$, thus
iterations diverge. When $0<\Delta t<2$, iterations converge, and
fastest convergence occurs when $\Delta t \approx 1.51$, which is
the value from Eq. (\ref{DT*AITEM}).

In the above numerical example, it is observed that $c_{opt}=\mu$.
Is this a coincidence? The answer is negative. Below, we will show
that for a large class of equations (\ref{ODE}) with localized
potentials, $c_{opt}=\mu$. This result is stated in the following
theorem.
\begin{theorem} \label{theoremAITEM}
Consider Eq.~(\ref{ODE}) with $\lim_{\bf |x| \To \infty} F(0, {\bf
x})=0$. If ${\cal V}({\mathbf x})$ given in (\ref{Vdef}) does not
change sign, then $c_{opt}=\mu$ in the AITEM
(\ref{t1_09a})-(\ref{t1_10}). If ${\cal V}({\mathbf x})$ changes
sign, then $c=\mu$ is not optimal in the generic case.
\end{theorem}
The generic case will be defined later in Lemma \ref{cmu}. It is
noted that when $\lim_{\bf |x| \To \infty} F(0, {\bf x})=0$, $\mu>0$
for solitary waves in Eq. ~(\ref{ODE}).

To facilitate the proof of Theorem \ref{theoremAITEM}, we first establish a few lemmas.

\begin{lemma} \label{lemmalambda}
If $\lim_{\bf |x| \To \infty} F(0, {\bf x})=0$ in Eq.~(\ref{ODE}),
then the continuous spectrum of operator ${\cal L} $ in
(\ref{operator}) is given by
\be \label{Lam_con}
\ba{ll}
\Lambda \in (-1, -\frac{\mu}{c}], & {\rm for}\;\; c>\mu; \vspace{0.5cm} \\
\Lambda \in [-\frac{\mu}{c}, -1), & {\rm for} \;\; c<\mu.
\ea
\ee
Moreover,  if $(\Lambda, \Psi)$ is a discrete eigenmode of Eq.
(\ref{operator}), then
\begin{equation} \label{Lambdaprime}
0<-\frac{1}{\Lambda}\frac{d\Lambda}{dc}=
\frac{\langle \Psi, \Psi \rangle }{\langle \Psi, M\Psi \rangle } < \frac{1}{c}.
\end{equation}
\end{lemma}

{\bf Proof:} \  First, since $\cL \to M^{-1}L_1$ as $\x \to \infty$,
the continuous spectrum of $\cL$ is then the same as that of
$M^{-1}L_1$, which can be easily shown to be (\ref{Lam_con}). Next,
we consider how a discrete eigenvalue $\Lambda$ changes with $c$.
The eigenvalue equation $\cL\Psi=\Lambda\Psi$ can be rewritten as
\begin{equation}
\label{LPsi}
L_1\Psi - H\hspace{0.03cm} u =\Lambda M \Psi,
\end{equation}
%
where $H=\langle L_1\Psi, M^{-1}u\rangle /\langle u,
M^{-1}u\rangle$. Differentiating Eq. (\ref{LPsi}) with respect to
$c$, then taking its inner product with $\Psi$ and noticing $\Psi
\in S$, we get
\begin{equation}
\langle (L_1-\Lambda M)\frac{\partial\Psi}{\partial c}, \Psi\rangle
= \langle \frac{d\Lambda}{dc} M\Psi + \Lambda \Psi, \Psi\rangle.
\label{t07_02}
\end{equation}
Since both $L_1$ and $M$ are self-adjoint and utilizing Eq.
(\ref{LPsi}), we have
\begin{equation}
\langle (L_1-\Lambda M)\frac{\partial\Psi}{\partial c}, \Psi \rangle
= \langle \frac{\partial\Psi}{\partial c}, (L_1-\Lambda M)\Psi
\rangle = H \langle \frac{\partial\Psi}{\partial c}, u \rangle=H
\frac{d}{d c}\langle \Psi, u \rangle, \label{t07_03}
\end{equation}
which is zero since $\Psi\in S$. Thus from Eq. (\ref{t07_02}), we
get
\begin{equation}
-\frac{1}{\Lambda}\frac{d\Lambda}{dc} = \frac{\langle \Psi, \Psi
\rangle }{\langle \Psi, M\Psi \rangle }. \label{dLdc}
\end{equation}
Since $\langle \Psi, M\Psi \rangle > c \langle \Psi, \Psi
\rangle>0$, Lemma \ref{lemmalambda} is proved. $\Box$

\begin{lemma} \label{cmu}
Consider Eq.~(\ref{ODE}) with $\lim_{\bf |x| \To \infty} F(0, {\bf
x})=0$. If either $\Lambda_{min}(\mu)=-1$ or
$\Lambda_{max}(\mu)=-1$, then $c_{opt}=\mu$. If neither
$\Lambda_{max}(\mu)$ nor $\Lambda_{min}(\mu)$ equals $-1$, then in
the generic case where
\begin{equation} \label{lamneq}
\left. \frac{1}{\Lambda_{max}}\frac{d\Lambda_{max}}{dc}\right|_{c=\mu} \ne
\left. \frac{1}{\Lambda_{min}}\frac{d\Lambda_{min}}{dc}\right|_{c=\mu},
\end{equation}
$c=\mu$ is not optimal.
\end{lemma}

{\bf Proof:} \  Differentiating formula (\ref{R*AITEM}) of $R_*(c)$
with respect to $c$, one gets:
\begin{equation}
\label{dRdc}
\frac{dR_*}{dc}=\frac{2\Lambda_{min}\Lambda_{max}}{(\Lambda_{min}+\Lambda_{max})^2}
\left[\frac{1}{\Lambda_{min}}\frac{d\Lambda_{min}}{dc}-
\frac{1}{\Lambda_{max}}\frac{d\Lambda_{max}}{dc}\right].
\end{equation}
The factor in front of the square brackets above is positive since
both $\Lambda_{max}$ and $\Lambda_{min}$ are negative (see beginning
of this section). Following the assumption of this lemma, suppose,
for definiteness, that $\Lambda_{min}(\mu)=-1$. When $c$ decreases
from $\mu$, the lower edge of the continuous spectrum decreases as
$-\mu/c$ [see Eq. (\ref{Lam_con})]. Due to inequality
(\ref{Lambdaprime}), all discrete eigenvalues of $\cL$ decrease
slower than $-\mu/c$, thus $\Lambda_{min}(c)=-\mu/c$. Then using
Eqs. (\ref{Lambdaprime}) and (\ref{dRdc}) one obtains:
\be
\frac{dR_*}{dc} < \left| \frac{2\Lambda_{min}\Lambda_{max}}{(\Lambda_{min}+\Lambda_{max})^2} \right|
\, \left[ -\frac{c}{\mu}\,\frac{\mu}{c^2} + \frac1c \right] = 0, \quad c < \mu.
\label{t07_04}
\ee
When $c$ increases from $\mu$, the lower edge of the continuous
spectrum is always at $-1$ [see Eq. (\ref{Lam_con})], while discrete
eigenvalues of $\cL$ all increase [see Eq. (\ref{Lambdaprime})] ,
thus $\Lambda_{min}(c)=-1$. Then using Eqs. (\ref{Lambdaprime}) and
(\ref{dRdc}) we get
\be
\frac{dR_*}{dc} > \left| \frac{2\Lambda_{min}\Lambda_{max}}{(\Lambda_{min}+\Lambda_{max})^2} \right|
\, \left[ 0 + 0 \right] = 0, \quad c > \mu.
\label{t07_05}
\ee
Inequalities (\ref{t07_04})--(\ref{t07_05}) mean that $R_*(c)$ has a
global minimum at $c=\mu$, thus $c_{opt}=\mu$. If
$\Lambda_{max}(\mu)=-1$, following similar arguments one can show
that $c_{opt}=\mu$ as well.

If, however, $\Lambda_{min}(\mu) <-1$ and $\Lambda_{max} (\mu) >-1$,
then under the condition (\ref{lamneq}) which holds in the generic
case, $R_*'(\mu)$ exists and is not equal to zero. Thus the minimum
of $R_*(c)$ is not at $c=\mu$, i.e. $c=\mu$ is not optimal. Lemma
\ref{cmu} is thus proved. $\Box$

In the following two lemmas, we establish the conditions under which
either $\Lambda_{min}(\mu)=-1$ or $\Lambda_{max}(\mu)=-1$. For
convenience, we define $M_0 \equiv M_{c=\mu} = \mu-\nabla^2$, ${\cal
L} _0 \equiv {\cal L}_ {c=\mu}$. Then $\Lambda_{min}(\mu)$ and
$\Lambda_{max}(\mu)$ are the smallest and largest eigenvalues of
${\cal L} _0$.

\begin{lemma} \label{equiv}
Suppose $\lim_{\bf |x| \To \infty} F(0, {\bf x})=0$ in
Eq.~(\ref{ODE}), then ${\cal L} _0$ and $M_0^{-1}L_1$ both do not
have continuous spectrum, and their discrete eigenvalues accumulate
at $-1$. In addition, if the smallest eigenvalue $\lambda_{min}$ of
$M_0^{-1}L_1$ is $-1$, then $\Lambda_{min}(\mu)=-1$; if
$M_0^{-1}L_1$ has two or more eigenvalues that are less than $-1$,
then $\Lambda_{min}(\mu) < -1$. Similarly, if the largest eigenvalue
$\lambda_{max}$ of $M_0^{-1}L_1$ is $-1$, then
$\Lambda_{max}(\mu)=-1$; if $M_0^{-1}L_1$ has two or more
eigenvalues greater than $-1$, then $\Lambda_{max}(\mu) > -1$.
\end{lemma}

{\bf Proof:} \ Operators $M^{-1}_0L_1$ and ${\cal L} _0$ do not have
continuous eigenvalues, as follows from Eq. (\ref{Lam_con}). The
eigenvalue equation $M^{-1}_0L_1\psi=\lambda \psi$ is the same as
the Schr\"odinger equation (\ref{sch2}). Using well known spectral
properties of the Schr\"odinger operators, we know that operator
$M^{-1}_0L_1$ has an infinite number of (discrete) eigenvalues,
which accumulate in such a way that $(1+\lambda)^{-1}$ approaches
either $+\infty$ or $-\infty$, or both (this fact for the 1D case
can be found in \cite{Ince}). Thus the accumulation point of
eigenvalues for $M^{-1}_0L_1$ is $-1$.

Applying the technique in the proof of Theorem \ref{theorem1} (see
Eqs. (\ref{Psihat})-(\ref{Q}), or \cite{Kivshar_book,VK}) on the
eigenvalue problem for $\cL_0$, one can show that between every two
adjacent eigenvalues of $M_0^{-1}L_1$, there is an eigenvalue of
${\cal L} _0$. Thus, $-1$ is also an accumulation point of
eigenvalues for ${\cal L} _0$. By the same reason, if the smallest
eigenvalue $\lambda_{min}$ for $M_0^{-1}L_1$ is $-1$, then
$\Lambda_{min}(\mu)=-1$; if $M_0^{-1}L_1$ has two or more
eigenvalues that are less than $-1$, then $\Lambda_{min}(\mu) < -1$.
Similarly, if the largest eigenvalue $\lambda_{max}$ for
$M_0^{-1}L_1$ is $-1$, then $\Lambda_{max}(\mu)=-1$; if
$M_0^{-1}L_1$ has two or more eigenvalues greater than $-1$, then
$\Lambda_{max}(\mu) > -1$. Lemma \ref{equiv} is thus proved. $\Box$

\begin{lemma} \label{lemmaMV}
Consider Eq.~(\ref{ODE}) with $\lim_{\bf |x| \To \infty} F(0, {\bf
x})=0$. For operator $M_0^{-1}L_1$, if ${\cal V}({\mathbf x})\ge 0$
for all ${\mathbf x}$, then its smallest eigenvalue $\lambda_{min}$
is $-1$; if ${\cal V}({\mathbf x})\le 0$ for all ${\mathbf x}$, then
its largest eigenvalue $\lambda_{max}$ is $-1$; if ${\cal
V}({\mathbf x})$ changes sign, then there is an infinite number of
its discrete eigenvalues on both sides of $-1$.
\end{lemma}

{\bf Proof:} \  Consider the eigenvalue equation
$M^{-1}_0L_1\psi=\lambda \psi$, which is the same as Eq.
(\ref{sch2}). Taking the inner product of Eq. (\ref{sch2}) with
$\psi$, we get
\begin{equation}
1+\lambda = \frac{\langle {\cal V}\psi, \psi \rangle}{\langle M_0
\psi, \psi \rangle}.
\end{equation}
If ${\cal V}({\mathbf x})\ge 0$ for all ${\mathbf x}$, then since
$M_0$ is a positive definite self-adjoint operator, the right hand
side of the above equation is non-negative, thus $\lambda > -1$. Due
to Lemma \ref{equiv}, eigenvalues of $M_0^{-1}L_1$ accumulate at
$-1$, thus $\lambda_{min}=-1$. By similar arguments, if ${\cal
V}({\mathbf x})\le 0$ for all ${\mathbf x}$, then
$\lambda_{max}=-1$. If ${\cal V}({\mathbf x})$ changes sign, then
Eq. (\ref{sch2}) has infinite numbers of discrete eigenvalues on
both sides of $-1$ \cite{Ince}. Hence Lemma \ref{lemmaMV} is proved.
$\Box$

With these lemmas, we are now ready to prove Theorem \ref{theoremAITEM}.

{\bf Proof of Theorem \ref{theoremAITEM}:} \ If ${\cal V}({\mathbf
x})$ does not change sign, then by Lemmas 4 and 5,
$\Lambda_{min}(\mu)=-1$ or $\Lambda_{max}(\mu)=-1$, hence by Lemma
3, $c_{opt}=\mu$. On the other hand, if ${\cal V}({\mathbf x})$
changes sign, Lemma \ref{lemmaMV} indicates that there are
infinitely many eigenvalues of $M_0^{-1}L_1$ both below and above
$-1$. Then by Lemma \ref{equiv}, there are also infinitely many
eigenvalues of ${\cal L}_0$ both below and above $-1$, so that
$\Lambda_{min}(\mu)<-1$ and $\Lambda_{max}(\mu)>-1$. Then in the
generic case defined by equation (\ref{lamneq}),
$c=\mu$ is not optimal by Lemma 3. Theorem \ref{theoremAITEM} is
thus proved. $\Box$

In practical implementations of the AITEM, after $c_{opt}=\mu$ in
Eq. (\ref{M}) is chosen, one still needs to select the time step
$\Delta t$. The best choice of $\Delta t$ which leads to fastest
convergence is given by Eq. (\ref{DT*AITEM}). Since the exact values
of $\Lambda_{min}$ and $\Lambda_{max}$ are usually not available,
below we give the interval of values where the optimal time step can
be found. When ${\cal V}({\mathbf x})\ge 0$ for all ${\mathbf x}$,
$\Lambda_{min}(\mu)=-1$, and $-1 < \Lambda_{max}(\mu) <0 $, hence
$1<\Delta t_*(\mu)< 2$. When ${\cal V}({\mathbf x})\le 0$ for all
${\mathbf x}$, $\Lambda_{min}(\mu) < -1 $ and
$\Lambda_{max}(\mu)=-1$, hence $\Delta t_*(\mu) < 1$.

In some physical problems, the assumption $\lim_{\bf |x| \To \infty}
F(0, {\bf x})=0$ is not met, i.e. the potential in Eq. (\ref{ODE})
is not localized. One example is the following type of equations
\begin{equation} \label{ODEnonlocal}
\nabla^2 u + V({\mathbf x}) u+ G(u^2, {\mathbf x}) u =\mu u,
\end{equation}
where $V({\mathbf x})$ is a periodic function in $\mathbf x$, and
$G(0, {\mathbf x}) \to 0$ as $\mathbf x \to \infty$. For these
equations, if we take the acceleration operator in the form of
$M=c-\nabla^2-V({\mathbf x})$, then we can also show that under
conditions analogous to those in Theorem 2, $c_{opt}=\mu$. However,
for this form of $M$, it is not easy to compute $M^{-1}$. So in
practice, it may still be better to use the simple form (\ref{M})
for Eq. (\ref{ODEnonlocal}) instead. In that case, $c_{opt}$ is not
known analytically and may need to be estimated by trial and error
(see Example 4 in Sec. \ref{examples}). Our experience shows that in
many cases, taking a suboptimal $c$ does not slow down the method
significantly as long as the $c$ taken is not far away from
$c_{opt}$. Thus using the form (\ref{M}) of $M$ for Eq.
(\ref{ODEnonlocal}) does not constitute a significant disadvantage.

Lastly, we would like to point out that most of the results in Secs.
3, 4 and 5 can be extended to a wider class of equations
\begin{equation} \label{generalODE}
{\cal D}u+F(u^2, {\mathbf x})u=\mu u,
\end{equation}
where $u$ is a real-valued localized function, $F$ is also
real-valued, $\mu$ is a real parameter, and ${\cal D}$ is a general
linear self-adjoint semi-negative-definite constant-coefficient
pseudo-differential operator. The previous equation (\ref{ODE}) is a
special case of (\ref{generalODE}) with ${\cal D}=\nabla^2$. For
this general equation (\ref{generalODE}), the AITEM is still
(\ref{t1_09a})-(\ref{t1_10}), with $\nabla^2$ replaced by ${\cal
D}$. The convergence conditions of this AITEM are still the same as
those in Theorem \ref{theorem1}, except that $\nabla^2$ in the
definition (\ref{L}) of $L_1$ is replaced by ${\cal D}$. Regarding
optimal acceleration, if $M=c-{\cal D}$ is taken, then we can also
show that $c_{opt}=\mu$ when $\lim_{|\x|\to \infty} F(0, \x)=0$ and
${\cal V}({\mathbf x})$ as defined in (\ref{Vdef}) does not change
sign. The connection between convergence of the AITEM and linear
stability of a nodeless solitary wave can also be extended to other
types of nonlinear wave equations. For instance, consider the
generalized one-dimensional Korteweg de Vries (KdV) equation
\be U_t+\left[{\cal D}U+F(U^2)U\right]_x=0, \label{KdV} \ee
where $U(x, t)$ and $F(\cdot)$ are both real-valued functions,
$F(0)=0$, and ${\cal D}$ is a general linear self-adjoint
semi-negative-definite constant-coefficient pseudo-differential
operator. Looking for moving solitary waves of the form $U(x,
t)=u(x-\mu t)$, where $\mu$ is a real parameter, we get an equation
for $u$ which is a special form of (\ref{generalODE}). For Eq.
(\ref{KdV}), it has been shown that if $u(x)$ has no nodes and
$p(L_1)=1$, then the solitary wave $u(x-\mu t)$ is linearly stable
if $P'(\mu)>0$ and linearly unstable if $P'(\mu)<0$ \cite{BSS87,
PW92}. In this case, for generic acceleration operators $M$ where
the eigenfunction of $M^{-1}L_1$'s positive eigenvalue is
non-orthogonal to $u(x)$, the convergence conditions of the AITEM in
Theorem \ref{theorem1} are the same as the stability conditions
above. Thus the connection between convergence of the AITEM and
linear stability of the solitary wave holds for Eq. (\ref{KdV}) as
well.

\section{The accelerated imaginary-time evolution method with amplitude normalization}

The AITEM (\ref{t1_09a})-(\ref{t1_10}) discussed above employed the
power normalization [see (\ref{t1_09a})], which has commonly been
used in all previous ITEM-type methods. A consequence of this
normalization is that when $P'(\mu)=0$, this method is doomed to
fail. In certain cases when $P'(\mu) <0$ (see case 2 in Theorem 1),
this AITEM diverges as well. In this section, we will propose a
different normalization for the AITEM which can overcome the above
difficulties.

This new normalization is the amplitude normalization. In other
words, instead of fixing the power of the solution we are seeking,
we fix the amplitude of $u(\x)$ (i.e., the largest value of
$|u(\x)|$). Thus, this new AITEM we propose for Eq. (\ref{ODE}) is
\begin{equation} \label{AN}
u_{n+1}=\frac{A}{|\hat{u}_{n+1}|_{max}} \hat{u}_{n+1},
\end{equation}
\be \hat{u}_{n+1}=u_n+ M^{-1}\left(L_{00}u-\mu u \right)_{u=u_n,\
\mu=\mu_n} \Delta t, \label{AN2} \ee
and
\be \mu_n=\left. \frac{\langle L_{00}u, M^{-1} u \rangle}{\langle u,
M^{-1} u \rangle} \right|_{u=u_n}, \label{AN3} \ee
where $A=|u|_{max}$ is the pre-specified amplitude of $u(\x)$. This
AITEM with amplitude normalization, which we denote as AITEM(A.N.),
can converge regardless of the value of $P'(\mu)$.

We have tested this AITEM(A.N.) on various examples, and found that
it is almost always more superior than the first AITEM
(\ref{t1_09a})-(\ref{t1_10}). This superiority is reflected on two
aspects: (i) in cases where the first AITEM does not converge, the
AITEM (A.N.) often can converge; (ii) in cases where the first AITEM
converges, the AITEM(A.N.) often converges faster.

To illustrate the first aspect of the AITEM (A.N.) superiority, we
consider the following example.

\begin{figure}[h]
\begin{center}
\parbox{12cm}{\postscript{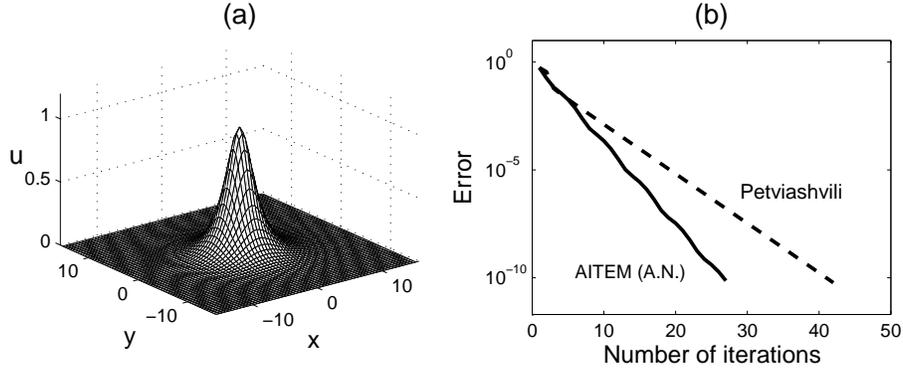}{1.0}}

\caption{(a) A solitary wave in the two-dimensional NLS equation
(\ref{NLS_2D}) with amplitude one; (b) error diagrams of the AITEM
(A.N.) and the Petviashvili method for this solitary wave.
\label{2DNLS_fig} }
\end{center}
\end{figure}

{\bf Example 3.} The 2D NLS equation
\begin{equation} \label{NLS_2D}
u_{xx}+u_{yy}+u^3=\mu u
\end{equation}
admits a family of single-hump (fundamental) solitary waves, all of
which have the same power $P=11.70$. Since $P'(\mu)=0$ everywhere,
the first AITEM (\ref{t1_09a})-(\ref{t1_10}) clearly can not
converge to these solitary waves. However, the AITEM (A.N.) can not
only converge, but also converge very fast. To illustrate, we select
the solitary wave with amplitude $A=1$, whose corresponding
propagation constant is $\mu=0.2054$. This solitary wave is
displayed in Fig. \ref{2DNLS_fig}(a). We take the acceleration
operator $M$ as (\ref{M}), the initial guess as a Gaussian hump
$u_0(x, y)=e^{-(x^2+y^2)/10}$, the spatial domain as a square with
side length of $30$, with each dimension discretized by 128 points.
We also take $(c, \Delta t)=(0.5, 1)$, which is nearly optimal. With
these choices, the AITEM (A.N.) rapidly converges to the solitary
wave [see the error diagram in Fig. \ref{2DNLS_fig}(b)]. Indeed, it
takes only 27 iterations for the error to fall below $10^{-10}$. For
comparison, we applied the standard Petviashvili method to this
example \cite{Pet}, taking the same spatial discretizations and
initial condition as above. The error diagram of the Petviashvili
method is also displayed in Fig. \ref{2DNLS_fig}(b). We see that the
AITEM (A.N.) is faster than the Petviashvili method by about 40\%.

The second aspect of the AITEM (A.N.) superiority over the first
AITEM will be illustrated by examples in the next section.

It should be pointed out that for the AITEM (A.N.), the connection
between convergence of the scheme and linear stability of the
solitary wave disappears. Regarding its convergence conditions, this
question can be analyzed by techniques similar to those used in
Secs. 3 and 5. This will not be done in this article, and will be
left for future studies.


\section{Examples illustrating convergence rates of the AITEMs}
\label{examples}

In this section, we will apply the two proposed AITEMs
(\ref{t1_09a})-(\ref{t1_10}) and (\ref{AN})-(\ref{AN3}) to two
physical examples, and compare their convergence speeds. In
addition, we will compare them to the Petviashvili method whenever
applicable.

\begin{figure}[h]
\begin{center}
\parbox{12cm}{\postscript{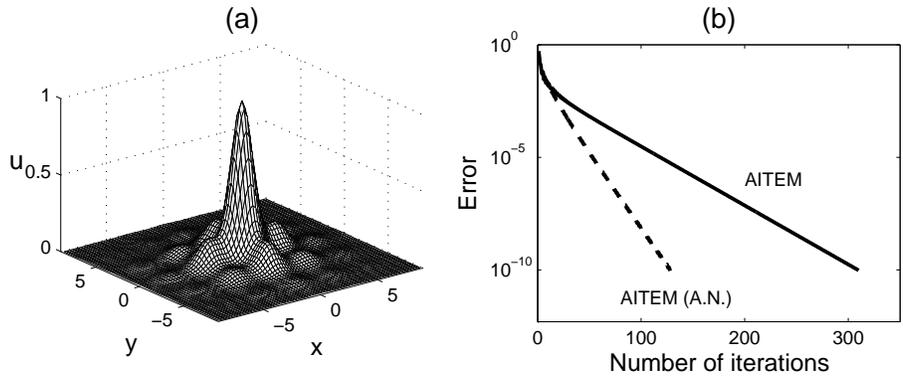}{1.0}}

\caption{(a) A solitary wave in Eq. (\ref{NLS_2D_latt}) with $V_0=3$
and $P=3$. (b) Error diagrams of the AITEM and AITEM (A.N.) [both
with $c=0.7$, $\Delta t=1$] for this solitary wave.
\label{fig_NLS_2D_lattice} }
\end{center}
\end{figure}

\noi {\bf Example 4} \ Let us first consider a 2D NLS equation with
a periodic potential,
\begin{equation} \label{NLS_2D_latt}
u_{xx}+u_{yy}+V_0 \left(\cos^2x+\cos^2y\right)u + u^{3}=\mu u,
\end{equation}
which has recently attracted much interest due to its application to
optical lattices and Bose-Einstein condensates
\cite{YangMuss03,Kivshar_BEC04,Christodoulides03}. This equation
admits a family of nodeless solitary waves. One of them with
$V_0=3$, $P=3$, $\mu=3.7045$ and amplitude $A=1.0384$ is displayed
in Fig. \ref{fig_NLS_2D_lattice} (a). For this wave, $p(L_1)=1$ and
$P'(\mu)>0$, thus the AITEM converges for generic choices of $M$. We
applied the AITEM and AITEM(A.N.) to search for this solitary wave.
The acceleration operator $M$ was taken as (\ref{M}), the spatial
domain taken to be a square with side length of $10\pi$, with each
dimension discretized by 128 points. As the initial guess, we took a
Gaussian profile $u_0(x, y)=\exp(-x^2-y^2)$. For our choice of
(\ref{M}), $c_{opt}$ is not known analytically. Hence we scanned
values of $c$ and $\D t$ in the AITEMs and found that taking $c=0.7$
and $\D t=1$ yielded the fastest (or nearly fastest) convergence for
both AITEM and AITEM(A.N.). The error diagrams versus the iteration
number for these two methods are shown in Fig.
\ref{fig_NLS_2D_lattice}(b). We see that an error below $10^{-10}$
is reached by the AITEM and AITEM(A.N.) in about 310 and 130
iterations respectively. Thus the AITEM (A.N.) converges much faster
than the AITEM. In the appendix, the matlab code of AITEM(A.N.) on
this example is attached, so that the reader can test this method
themselves. For Eq. (\ref{NLS_2D_latt}), the original Petviashvili
method does not apply \cite{Pet}. Several generalizations of that
method for equations such as (\ref{NLS_2D_latt}) have been proposed
recently \cite{Ablowitz_Muss,LakobaYang,MussYangJOSAB}. A comparison
between those generalized Petviashvili methods and the AITEMs in
this paper will be considered elsewhere.

\begin{figure}[h]
\begin{center}
\parbox{12cm}{\postscript{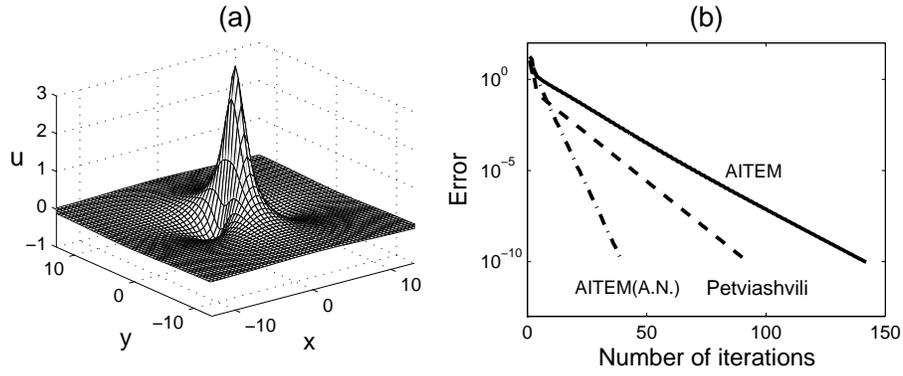}{1.0}}

\caption{(a) A solitary wave in the KP equation (\ref{KP}) with
$\mu=1$; (b) error diagrams of the AITEM (with $c=1.4, \Delta
t=1.7$), AITEM (A.N.) (with $c=1.4, \Delta t=1.4$) and the
Petviashvili method for this solitary wave. \label{fig_t2_KP} }
\end{center}
\end{figure}

\noi {\bf Example 5} \ Another example we consider is an equation
whose linear part is not of the Schr\"odinger type. Specifically, we
consider the integrable Kadomtsev-Petviashvili (KP) equation, whose
stationary solutions satisfy the equation
\be
u_{xx}-\partial^{-2}_x u_{yy}+u^n=\mu u, \qquad n=2
\label{KP}
\ee
and the constraint
\be
\int_{-\infty}^{\infty} u(x,y) dy =0\,.
\label{c_KP}
\ee
The analytical expression for solitary waves of Eq. (\ref{KP}) is
\cite{ManakovZ_77}
\be
u=12\mu\, \frac{3+\mu^2y^2-\mu x^2}{(3+\mu^2y^2+\mu x^2)^2}\,,
\label{sol_KP}
\ee
and their power function is $P(\mu)=24\pi\sqrt{\mu}$. It is known
that $p(L_1)=1$ for solutions (\ref{sol_KP}) \cite{Peli_Pet}, and
$P'(\mu) >0$ in view of the above power formula. Thus the AITEM
converges to these localized solutions for generic choices of $M$
(see end of Sec. \ref{sec_opt}). We applied the AITEM, AITEM (A.N.)
and the Petviashvili method \cite{Pet} to compute one of these
solutions with $\mu=1$, whose profile is shown in Fig.
\ref{fig_t2_KP}(a). The spatial domain was taken as a square with
side length of 120, discretized by $512$ points in each dimension.
The acceleration operator $M$ in the AITEMs was taken as
$M=c-\partial_{xx}+\partial_x^{-2}\partial_{yy}$. Since ${\cal V}(x,
y)=2u$ changes sign here, generically $c_{opt}\ne \mu$ in the AITEM.
For the AITEM and AITEM (A.N.), we took $(c, \Delta t)=(1.4, 1.7)$
and $(1.4, 1.4)$ respectively, which yield near-optimal convergence
for the underlying methods. For all three methods, we imposed the
constraint (\ref{c_KP}) by setting the coefficient of the ${\mathbf
k}_x=0$ Fourier harmonic to be zero at every iteration. The error
diagrams for these methods are displayed in Fig. \ref{fig_t2_KP}(b).
This time, an error below $10^{-10}$ is reached by the AITEM,
AITEM(A.N.) and the Petviashvili method in about 140, 40 and 90
iterations respectively. Thus the AITEM (A.N.) converges much faster
than the Petviashvili method, while the Petviashvili method
converges faster than the AITEM.

We have studied solitary waves in the generalized KP equation
(\ref{KP}) (with $n\ne 2$) as well. In those equations, if $n>7/3$,
solitary waves are unstable \cite{WangA_94}. In such cases, we found
that the AITEM (\ref{t1_09a})-(\ref{t1_10}) diverged. Thus the
connection between convergence of that AITEM and the stability of
the underlying solitary wave holds for the generalized KP equation
as well.

\smallskip

From the above examples, we conclude that the AITEM (A.N.) converges
faster than the AITEM and Petviashvili methods, while the AITEM and
the Petviashvili method are comparable in convergence speeds.

Before concluding the paper, we make a comment on the practical
implementations of the AITEMs (and other methods such as the
Petviashvili method as well). If the convergence theorem predicts
that a method diverges for a solitary wave, sometimes iterations can
still converge (at least to a certain accuracy). This can happen,
for instance, when the iteration operator ${\cal L}$ has a single
unstable symmetric eigenmode, but the initial condition is chosen to
be, and the final solution is, strictly anti-symmetric. In that
case, the unstable symmetric eigenmode may not yet be excited before
the iterations have already converged. Such an example is Eq.
(\ref{1Du_double}), which admits anti-symmetric solitary waves. When
$P=3$ (see Fig. \ref{nodes_soliton}), $p(L_1)=1$, and $P'(3)<0$.
Hence the AITEM (\ref{t1_09a})-(\ref{t1_10}) should diverge.
However, we found that if we used anti-symmetric initial conditions,
then the AITEM iterations can converge to the solution with error
below $10^{-12}$. However, if a small symmetric component was
introduced into the initial condition, the iterations would diverge.
Thus, the convergence results obtained in this paper must be
understood as pertaining to generic initial conditions.  If
non-generic initial conditions are used in practical implementations
of the AITEMs, better convergence behavior may be observed.

\section{Summary}

In this paper, we proposed two accelerated imaginary-time evolution
methods for computations of solitary waves in arbitrary spatial
dimensions. The first method is the AITEM
(\ref{t1_09a})-(\ref{t1_10}) with the conventional power
normalization. For this method, the convergence conditions were
derived. These conditions show that this method usually converges to
nodeless solutions, but there also exist cases when this AITEM
converges to solitary waves with nodes. For nodeless solutions, we
also showed that this AITEM converges if and only if the solitary
wave is linearly stable. Conditions for optimal accelerations of
this AITEM were also derived. The second method we proposed is the
AITEM (A.N.) (\ref{AN})-(\ref{AN3}) which employs a novel amplitude
normalization. Both methods were applied to several examples of
physical interest, and we found that the AITEM (A.N.) delivers the
best performance, while the AITEM and the Petviashvili method are
comparable in performance.

\section*{Acknowledgment}
The work of J.Y. is supported in part by The Air Force Office of
Scientific Research under grant USAF 9550-05-1-0379, and the work of
T.I.L. is supported in part by the National Science Foundation under
grant DMS-0507429.

\section*{Appendix: \ Matlab code of AITEM (A.N.) for Example 4}

\begin{verbatim}
Lx=10*pi;  Ly=10*pi;  N=128;  c=0.7;  DT=1;
max_iteration=10000; error_tolerance=1e-10;

dx=Lx/N;  x=-Lx/2:dx:Lx/2-dx;  kx=[0:N/2-1 -N/2:-1]*2*pi/Lx;
dy=Ly/N;  y=-Ly/2:dy:Ly/2-dy;  ky=[0:N/2-1 -N/2:-1]*2*pi/Ly;
[X,Y]=meshgrid(x, y);  [KX,KY]=meshgrid(kx, ky);

A=1.0384;
U=exp(-(X.^2+Y.^2));  U=U/max(max(abs(U)))*A;
for nn=1:max_iteration
    Uold=U;
    L00U=ifft2(-(KX.^2+KY.^2).*fft2(U))+3*(cos(X).^2+cos(Y).^2).*U + U.^3;
    MinvU=ifft2(fft2(U)./(c+KX.^2+KY.^2));
    mu=sum(sum(L00U.*MinvU))/sum(sum(U.*MinvU));
    U=U+ifft2(fft2(L00U-mu*U)./(KX.^2+KY.^2+c))*DT;
    U=U/max(max(abs(U)))*A;
    Uerror(nn)=sqrt(sum(sum(abs(U-Uold).^2))*dx*dy); Uerror(nn)
    if Uerror(nn) < error_tolerance
        break
    end
end
\end{verbatim}

This code as well as Matlab codes for other examples can be
downloaded at \\ www.cems.uvm.edu/\verb+~+jyang/Publication.htm

\newpage

\end{document}